# THE ADOPTION OF ROBOTICS BY GOVERNMENT AGENCIES: EVIDENCE FROM CRIME LABS

Andrew B. Whitford*
University of Georgia

Jeff Yates
Binghamton University

Adam Burchfield
University of Georgia

L. Jason Anastasopoulos
University of Georgia

Derrick M. Anderson
Arizona State University

## Abstract

The adoption and use of emerging and smart technologies like robotics is a key concern as the advanced economies undergo the latest technological revolution. While firms and factories often adopt technologies like robotics and advanced manufacturing techniques at a fast rate, government agencies are often seen as lagging in their adoption of such tools. We offer evidence about the adoption of robotics from the case of American crime laboratories. Using data from the census of crime labs, we show that the adoption of such technologies depends on traditional drivers of agency capacity and demand: budgets, the agency's task environment, and its relative level of professionalism. Together these findings suggest that agencies can be early adopters of such technologies if they have the capacity (and need) to do so.

* Corresponding author.

**INTRODUCTION**

In their 2014 book, Erik Brynjolfsson and Andrew McAfee argue that human civilization is at the beginning of a "second machine age" - an age in which machines and algorithms will replace skilled and cognitive work. The age of robots and artificial intelligence will substitute, instead of complement, humans. The work we do and the way organizations work will both change.

Klaus Schwab makes similar arguments in his 2017 book "The Fourth Industrial Revolution". In that framework, human work is augmented by machines, algorithms, and even more esoteric technologies (like easy DNA sequencing). These changes offer opportunities - and risks. For Schwab, the goals are to empower people rather than to replace, to serve rather than disrupt, and to maintain the ethical boundaries that have underpinned much of human existence.

Both views tell of a new world inhabited by machines of daunting capabilities. This is the world of "smart technology" that promises improvements and claims attention by prognosticators and technologists alike. For very good reasons, the adoption and use of emerging and smart technologies like robotics is a key concern in advanced economies. But technological changes do not occur in vacuums. The world's organizations make decisions about interacting with such technological shifts, so organizations can act as the conduits of technological change.

While firms and factories often adopt technologies like robotics and advanced manufacturing techniques at a fast rate, government agencies are often seen as lagging in their adoption of such tools. Notwithstanding sci-fi narratives about the use of scary machines by government agencies, there are just as many narratives about broken air conditioners and eternal lines. Public access to the Internet is now 30 years old, and governments large and small still struggle with maintaining usable web pages.

Such narratives are convenient but likely a distraction. This paper departs from such narratives to offer two kinds of evidence about the advent of smart technology - in the form of robotics - in government service. First, we offer evidence about the adoption of robotics from the case of American crime laboratories. The crime labs play central roles in the American criminal justice system as adjudicators of evidence - a role that is closely watched in a world where police and prosecutors face their own unique incentives. Their pivotal position (and the demands placed on them) is made clear from current political debates about the massive backlogs in processing rape kits.

Second, using data from two censuses of the crime labs, we investigate possible drivers of the adoption of the robotics as smart technology. We offer hypotheses about that adoption process drawn from the broader literature on technology adoption in government. We note, though, that these hypotheses are rooted in very traditional concerns that underpin much of our received wisdom about government agencies. We show that the adoption of such technologies depends on traditional drivers of agency capacity and demand for innovation: budgets, the agency's task environment, and its relative level of professionalism.

Together these findings suggest that agencies can be early adopters of such technologies if they have the capacity (and need) to do so. Because organizations are the conduits for technological change, it is foreseeable that what makes organizations adopt robotics should be similar to what makes organizations do what they do most of the time. Yet, there are aspects of our findings that help us understand the pace of adoption of such smart technologies - about what may drive early or late adoption, and consequently what may underpin the inevitable geographic inequities that result from such processes. A second machine age or a fourth industrial revolution is probably inevitable; but as author William Gibson said in the 1990s, "The future is already here — it's just not very evenly distributed".[1]

This paper offers one view on the adoption of smart technologies like robotics by an array of public agencies, all within a single practice space. The next section provides a broad overview of the adoption of smart technologies, both by firms and within the public sector; we also introduce robotics as a special case of emerging (smart) technology. Section 3 describes our case environment - the crime laboratories - and what they can tell us about broader adoption processes in the public sector. The fourth section reviews our hypotheses, and the fifth section introduces the data and their attributes. After that, we describe our estimation approach and the results of our models. Finally, we return to the theme of the adoption of emerging smart technologies and what their adoption can tell us about the future of public organizations.

## CAPACITY AND THE ADOPTION OF EMERGING TECHNOLOGIES

Smart technologies (STs) may greatly enhance the ability of government organizations to effectively carry out their missions (Fredricksen and London 2000; Eisinger 2002). This increase in organizational capacity may come if STs help automate tasks typically requiring high degrees of expertise and specialization. Organizational theorists and practitioners have long known this. For instance, in public libraries sorting and cataloging books, journals, and newspapers in the past required specialized knowledge of the Dewey Decimal System and the collective, coordinated efforts of highly-trained, specialized librarians and information scientists. We know now that STs such as unsupervised machine learning (ML) algorithms like topic models can perform similar classification tasks for hundreds of thousands of documents, and do so with the help of only a single librarian with some knowledge of ML.

---

[1] https://www.npr.org/2018/10/22/1067220/the-science-in-science-fiction at 11:55.

Potential short-term costs of adopting STs may include additional infrastructure investments, employee training, and in some cases reorganization of workflow and workforce restructuring. Even so, the long-run gains in productivity and profitability that STs can bring have caused firms in the private sector firms to adopt them just on the basis of cold economic calculation.

In the public sector, however, the underlying causes of the adoption of such STs is less clear - especially in agencies not directly accountable to the public if the adoption of STs may threaten the careers of highly specialized employees. We focus here on a set of reasons why public agencies may want to increase their organizational capacity in a relatively rapid and significant way. We first discuss the types of agencies that can benefit most and least from adopting STs. We focus throughout on how those benefits can flow from characteristics that are familiar to many who study public sector organizations, such as the degree of public accountability, revenue streams, changing demand for services, and the amenability of important tasks to automation.

While most agency staff are unelected, the reputations, budgets, and fortunes of many agencies depend upon the efficient and timely delivery of services. For instance, the Food and Drug Administration (FDA), helps shape the fortunes of pharmaceutical manufacturers by conducting clinical trials and drug testing in a timely manner (Carpenter 2014; Carpenter 2002). Similarly, the U.S. Postal Service (USPS) is asked to deliver almost half a billion mailpieces per day, many of which are now parcels whose safe delivery could have serious consequences for the public and also for the agency's reputation.[2]

It is not surprising that agencies like the USPS have some of the most sophisticated and expansive information technology (IT) infrastructures. The USPS owns the third largest IT infrastructure in the world[3], having automated a large portion of their services. The agency has also participated in the ML, artificial intelligence, and big data revolutions during a time when few even knew about these technologies. Indeed, in the 1990s the USPS provided thousands of samples of handwritten digits to the computer scientists who used them to develop neural networks, currently one of the most powerful ML algorithms used to detect objects within images (see eg. LeCun et. al. 1990).

Because agencies recognize the need to respond (if only to protect their reputation), they must find ways to respond to changes in demand for their services. If demands for services increase dramatically, and if the agency's capacity is not sufficient, they have little choice but to find ways to increase their ability to meet demand without expecting significant budget increases. This is a path supported by STs. A high-capacity agency that faces increased demand may be able to better utilize their currently resources to meet increasing demand without resorting to automation-based approaches.

As an example, since 2010 the Internal Revenue Service (IRS) has experienced significant decreases in capacity, having lost an average of 16 percent of its funding and a whopping one-third of its workforce[4]. Such reductions in capacity have natural consequences, such as reducing its ability to enforce tax policy or conduct audits. While much has been written about the IRS's management and situation, it is notable that as a response to falling capacity and constant or increasing demand, the IRS has turned to bots to automate its vendor compliance process.[5] This tool cost about $190,000 to build but freed up thousands of staff hours for tasks other than vendor compliance.[6]

---

[2] https://facts.usps.com/one-day/
[3] https://about.usps.com/strategic-planning/cs09/CSPO_09_051.htm
[4] https://www.nextgov.com/emerging-tech/2019/04/irs-turns-automation-amid-shrinking-workforce/156161/

Of course, ST adoption requires that important tasks that consume significant hours be amenable to automation. This implies that labor-intensive tasks in the agency should have a relatively clear set of rules to inform the ST that is doing the task. Such tasks include sorting mail by zip code, region, or alphabetical order because these can be accomplished faster and better (with a lower error rate) than by a human.

Other tasks require complex reasoning and bureaucratic discretion. For instance, federal law sets certain criteria for granting asylum to refugees in the U.S., but those criteria are very broad and require holistic evaluation of a refugee's situation on a case-by-case basis that involves an interview process and background check. Costs might encourage the automation of such tasks but the work of USCIS officers is unlikely to be effectively automated by an ML algorithm or ST in the near future.

These vignettes suggest that while technology adoption in industrial organizations is often driven by the pursuit of market advantages, technology adoption by government is more likely to be associated with resource constraints, or a pursuit of public goals such as transparency, new or improved service delivery, or communication with citizens or stakeholders. The academic literature also documents these drivers (Bertot et al., 2010; Gil-Garcia, Helbig and Ojo, 2014; Reddick and Turner, 2012). The strategic adoption of technology can improve public engagement or access to information; it can improve decisions and operational efficiency (Desouza and Bhagwatwar, 2012). While most of the academic literature on technology adoption in agencies centers on information technologies, the findings there still help us understand how agencies consider and adopt STs.

---

[5] https://www.nextgov.com/emerging-tech/2019/04/irs-turns-automation-amid-shrinking-workforce/156161/
[6] https://www.nextgov.com/emerging-tech/2019/04/irs-turns-automation-amid-shrinking-workforce/156161/

For instance, much of the existing literature centers on technology adoption to improve participation in democratic governance - sometimes with only marginal effect, sometimes leading to more comprehensive transformation. The Internet has stimulated various "open government" and e-governance movements, and scholars observed the shift in governance "from broadcast-oriented models of information dissemination toward more social, citizen engagement focused models" as governments harness the Internet (Mascaro and Goggins 2011, 11). Consequently, participation shifts to digital or virtual interfacing, so not adopting such emerging technologies puts governments at a disadvantage (Desouza and Bhagwatwar, 2012).

Accordingly, concepts like "smart governance" see technology as part of a more holistic governance approach (Gil-Garcia, Helbig and Ojo, 2014). In the pursuit of flexible, open, participatory government, the adoption of new technologies must also come with "structural renewal"; this includes decentralization, adaptive management, and openness to collaboration with external stakeholders and communities (Kliksberg, 2000). In a pertinent example, regarding Baltimore's successful use of big data to improve government services, former Mayor Martin O'Malley called for governments to facilitate connection and collaboration across the full range of stakeholders (governors, technology managers, citizens, etc.) - activities that also may make them more likely to leverage new technology (O'Malley, 2014).

Adoption brings with it a need for new operational logics, protocols, rules and policies - if only because of the effects that new technology may have on communities (Barben et al., 2008). Oversight itself plays a role here if its implementation results in different structures for technology implementation (Dawson, Denford and Desouza, 2016). The balance between oversight and autonomy helps determine government innovation. Of course, removing the "shackles of oversight and administration" does not naturally lead to innovation (Dawson and Denford, 2015, 9). Those most successful at spurring innovation integrate oversight, supportive technology governance structure, clear goals and leadership involvement (Dawson, Denford and Desouza, 2016).

Yet, government agencies often have risk-averse cultures and especially so in the context of technology adoption (Bozeman and Bretschneider, 1986; Desouza, 2009). In a world where the consequences of failed implementation often overshadow or negate the benefits of innovation, public organizations may adopt new technologies only after a certain level of necessity is met or surpassed. In a study by Dawson and Denford (2015) of federal agency officials, many agencies were found to not enabled their employees to be innovative and have instead discouraged them from taking on high-risk, high-reward activities. Agencies may not adopt new technologies out of fear that the technological disruption will highlight shortcomings in specific programs or departments, potentially drawing criticism from management, elected officials, and the public (Dawes and Nelson, 1995). A shortage of IT talent can block technology adoption in local governments (Norris et al., 2001), or employees may resist change (Ebrahim and Irani, 2005; Li and Steveson, 2002), or senior managers may not recognize the strategic benefits of new technology (Fletcher and Wright, 1995).

But personnel factors are not the only barriers. Assorted institutional and political barriers have blocked e-government adoption, including limited communication and data sharing among agencies when centralization is needed in the implementation of IT-related initiatives (Aichholzer and Schmutzer, 2000; Ebrahim and Irani, 2005). Other limiting factors include concerns around information privacy and security (Ebrahim and Irani, 2005; Savoldelli, Codagnone and Misuraca, 2014). It is also difficult to evaluate the effectiveness of some new technologies (Ebrahim and Irani, 2005; Savoldelli, Codagnone and Misuraca, 2014).

Perhaps the biggest limiting factor is resources because many technology implementation initiatives require substantial infrastructure investments. Both the initial investment and maintenance costs can be prohibitive for smaller governments and public agencies (Ebrahim and Irani, 2005). Sporadic or inconsistent funding from a central budget source (e.g, a municipal general fund) - and especially so if it is subject to political influence (Heeks, 2001). This is especially true "if it is a tactical or strategic, rather than a more straightforward, operational innovation" (Dawson and Denford, 2015, 21).

The adoption of a complex smart technology by a government agency could bring great social benefits. As in so many other contexts, though, we do not know enough about the drivers of such decisions. On one hand, agencies have and will continue to adopt such technologies. On the other hand, many agencies remain mired in older patterns of producing public services. In this discussion, though, one thing should be kept in focus: that the technological landscape is changing faster than the government agencies charged with protecting the public and producing public services. The fact that we are still debating the conditions under which governments will use the Internet to deliver services should reveal the depth of the disconnect.

**The Robotics Revolution**

In this paper, we depart from previous papers that have sought to understand why and how governments use technologies like the Internet. Our focus is on technologies like those that make up what some call the coming "second machine age" (Brynjolfsson and McAfee 2014) or the "fourth industrial revolution" (Schwab 2017). This paper centers on the adoption of robotics by government agencies.

As is often the case with fast moving fields like emerging technologies, defining terms can be difficult. The field of robotics is "the science or the study of technology involved in the design, development and deployment of machines, known as robots, to perform tasks normally performed by human beings" (Kernaghan 2014, 488-89). More complicated is an easy definition of "robot" because some disagreement exists over the degree of autonomy necessary for a machine to be appropriately considered a robot (Calo 2015). However, a certain degree of consensus exists on core concepts and robots can perhaps be best thought of as "objects or systems that sense, process, and act upon the world to at least some degree" (2015, 531).

The concept of robots has been with us since at least ancient Greece (Kernaghan 2014), and arguably even earlier – depending on one's definition of the term. They began to come into their own as viable and useful devices in the mid-twentieth century with the advent and popularization of industrial robots, and by 2008 the world robot population was estimated to have reached approximately 8.6 million (2014, 489). By this time, the development and expanse of robots had followed a different narrative than many popular writers and Hollywood had promised in the 1950s. Rather than working as trusty servants or being nearly indistinguishable from humans, robots had been created and adapted to take on a number of highly technical, sometimes dangerous, and often mundane tasks – finding their primary role in business rather than the home (Moravec 2009).

In recent years, the world has witnessed a massive increase in the development and usage of robots (e.g. Sander and Wolfgang 2014; Calo 2015; Ford 2015; Cheng, Jia, Li, and Li 2019). Experts explain that this expansion has affected labor and production markets and will continue to do so, but to a much greater extent than in past decades. The manner, form, and dynamics of this impact on a given locale or industry (e.g. the United States, or electronics) is, in a word, complex (e.g. Sander and Wolfgang 2014; Ford 2015). The upward trajectory in robotics technology should not underestimated. One report projects that spending on robotics will leap from $15 billion in 2010 to $67 billion in 2025 (Sander and Wolfgang 2014). This dramatic growth has been especially pronounced in China. In 2000, China made up only about 0.4 percent of world total robot sales. By 2016, China accounted for approximately 30 percent of global sales and currently holds the world's largest operational stock of robots – about 19 percent of total world stock (Cheng, Jia, Li, and Li 2019, 73).

In the United States, robots have been used for a vast array of purposes. While their primary usage has been in industry and production, robots also have also found a home in service and other business, personal, and public applications. In the criminal justice sphere alone, robots have proven useful in a range of activities, including certain police practices, ballistics, and forensic crime analysis (primarily DNA testing). In recent years, some criminal justice units have successfully employed robots to address backlogs in rape kit testing (French 2017). Advances in robotics technology will bring about substantial legal, social and ethical challenges. Public service entities, such as criminal justice agencies, will need to carefully consider how to manage such technological advancements (Kernaghan 2014, 486).

Robots are the leading edge of the fourth industrial revolution, one that will also include the commonplace deployment of AI, wearable sensors and microchips, additive manufacturing, and easy DNA sequencing. Our paper provides a view into the everyday use of such technologies by regular government agencies. We are not focused here on the deployment of such technologies by military agencies in combat situations or by the Department of Energy for handling nuclear waste. As we discuss below, we focus on state, local, and regional agencies in the U.S. as a way of showing how the fourth industrial revolution is already here. The adoption and deployment of smart technology is already happening in many places.

**THE CASE OF THE CRIME LABS**

Our sample is drawn from a census of all of the operating crime laboratories in the U.S. Crime labs examine and report on physical evidence collected during criminal investigations for federal, state, and local jurisdictions. They perform a variety of forensic analyses and receive requests for these services from criminal justice agencies, such as police departments, prosecutors' offices, courts, and correctional facilities. Crime laboratories play a critical role in the justice system, analyzing millions of pieces of evidence from criminal investigations each year. In 2014, the nation's 409 crime labs received an estimated 3.8 million requests for forensic services. In 2014, more than half of the requests for forensic services received by crime labs nationwide were submitted to state labs (Durose, et al. 2016, 1).

The ability of a crime lab to handle its forensic workload depends on many factors, including the complexity of the procedures and the availability of resources. A criminal case may require more than one type of request to process or analyze evidence. For example, a crime lab may receive fingerprints and DNA evidence from the same case, which requires two separate analysis by different sections of the lab. About three-quarters of requests received in 2014 were for either analysis of controlled substances or biological samples collected for DNA profiles. Forensic biology casework accounted for a larger proportion of the overall number of requests received in 2014 than in 2009 (Durose, et al. 2016, 3). The composition of the forensic work handled varies among those serving federal, state, county, and municipal jurisdictions.[7] Thirty-eight percent of crime labs outsourced one or more types of forensic services during 2014; municipal labs were more likely than federal and state labs to outsource requests for services (Durose, et al. 2016, 4). In addition to their budgets, crime labs received funding from other sources, such as grants and fees.

While we discuss more attributes of the crime laboratories in detail below, we highlight two aspects of these organizations that make them a useful test case for considering the advent of robotics in public agencies. First, the crime labs face complex and shifting work environments. Rather than point to Hollywood depictions of crime scene investigators, we simply recount what they do every day: analyzing millions of pieces of evidence from criminal investigations each year. Rather than factories, they are more like academic laboratories. Second, the labs are populated by educated and skilled individuals. The tasks they face require using portfolios of abilities gained over long periods of time and at substantial cost. They are more like detectives than patrol officers. Again, rather than being like factory workers, lab staff are highly-specialized technicians.

---

[7] Toxicology requests accounted for 25 percent of the requests received by county labs, compared to less than 10 percent received by federal and municipal labs. In comparison, biological samples collected from convicted offenders and arrestees for DNA profiles comprised 39 percent of requests made to federal labs and 36 percent of requests to state labs, compared to less than 5 percent of requests to county and municipal labs. Crime scene investigations accounted for 17 percent of requests made to municipal labs, compared to 9 percent of requests to county labs and less than 1 percent of requests to state and federal labs. Crime labs provided an average of five different forensic functions in 2014. Eighty-one percent of crime labs handled the identification of illegal drugs and other controlled substances in 2014. Sixty-two percent of crime labs analyzed biological samples, such as blood and saliva, during 2014. Sixty-one percent analyzed forensic biology collected during criminal casework from crime scenes, victims, or suspects, and 16 percent analyzed biological samples collected from convicted offenders and arrestees for inclusion in a local, state, or national DNA database (Durose, et al. 2016, 2).

The consequence of these two aspects is that the laboratories provide a unique and compelling lens into how and when agencies adopt emerging technologies. The crime labs are adjudicators of evidence in the American criminal justice system, thus playing a pivotal role when police officers or prosecutors pursue their own incentives. But the labs are also a microcosm of government in the coming fourth industrial revolution. They are places where new and emerging technologies may or may not take hold - with attendant consequences for the overall performance of government.

**CONSTRAINTS ON INNOVATION IN AGENCIES**

Our goal is to estimate a function of the likelihood of adopting robotics by commonplace government agencies. In order to do so, in this section we offer hypotheses about how externally-observable aspects of agencies might be associated with that likelihood.

Of course, the adoption of technologies depends on macro-trends that work across institutional or environmental contexts. Yet, adoption by a specific group, institution, or industry will be driven by contextual forces, incentives, and orientation that affect and help define a given entity. In the case of publicly-funded forensic crime laboratories we consider these contextual influences; each lab has a distinct set of pressures and incentives - even if overlap exists among units. These laboratories live in different worlds that both enable and constrain the technologies they adopt and use – and that also affect the ultimate likelihood of their success.

As noted above, the rise of robotics has been understudied generally given how ubiquitous such technology has become in everyday life. This imbalance has been especially notable with regard to the relatively new upswing of robotics in government agencies. As such, our examination of adoption of robotics is fairly exploratory as a strong theoretical grounding on this has not yet developed. Of course, we are not starting completely from a blank slate. The policy innovation and adoption literature is well-developed and there exists a robust, albeit smaller, literature concerning government adoption of technological advancements. These include the adoption of computer technology (Brudney and Selden 1995); e-government (broadly) (Moon 2002; Norris and Moon 2005; Jun and Weare 2010; Monoharan 2012), social media (Mergel and Bretschneider 2013); and Geographic Information Systems (GIS) (Nedovic-Budic and Godschalk 1996), among others.

Research on innovation adoption focuses on the central question of why one organization is more likely to adopt an innovation, such as a new technology, than another organization (Brudney and Selden 1995, 71). Government units already deal with competing claims on their resources, energy, and time. The prospect of a new tax on an organization's resources and time is taken very seriously. In the case of robotics innovations, adoption involves costs of time and resources, and may also introduce employee confusion and new learning curves. Furthermore, adoption of new technologies such as robotics may even be disruptive to long-standing power structures and work routines within an organization (Jun and Weare 2010, 496). With this in mind, adoption of new technologies also presents the potential of gains along a number of fronts – efficiency, enhanced capabilities, and even prestige.

Our focus here is on organizational constraints and incentives. We draw from the more general literature on innovation adoption discussed above, and specifically technological adoption, to offer a set of core concepts that we believe help explain differences among laboratory choices to employ robotics technology. Our first potential explanation centers on unit resource capability and flexibility. As Holden, Norris and Fletcher (2003, 339) found in their survey of local government units, one of the foremost cited barriers to technological adoption or augmentation of technology is financial resources. Technological adoptions (such as robotics) can be costly, therefore we may reasonably expect that laboratories with more robust budgets are more apt to augment their facility and staff with robotics technology (Brudney and Selden 1995; Norris and Moon 2005; Manoharan 2012). We offer the following hypothesis:

H1: The probability of adoption increases when budgets are greater.

Adopting innovative technologies, such as robotics, can potentially help laboratories to deal with pressing concerns involving their task environment (e.g. Jun and Weare 2010). Government organizations can experience challenging task loads, both from internal and external sources. However, in this instance, we are interested primarily in the request loads that laboratories receive from external sources, specifically criminal justice agencies. Increasing requests require that a lab typically either hire more personnel or find another manner to deal with bigger workloads. Adoption of robotics technology can present a long-term cheaper and possibly more efficient approach to dealing with increases in external requests. We offer this hypothesis about the task environment:

H2: The probability of adoption increases when task demands are greater.

A principal theory in innovation adoption deals with the influence of professionalism in government organizations. An institutional environment that is more professional is typically associated with the efficient and effective delivery of government goods and services and the adoption of innovative technology often facilitates this aim (Brudney and Selden 1995; Jun and Weare 2010). With publicly-funded forensic crime labs there are two intuitive available measures for professionalism – accreditation and proficiency testing. The majority of forensic crime labs examined were accredited and employed proficiency testing. Accreditation is voluntary and there are a number of accrediting professional organizations for forensic crime labs – the most popular accrediting body is the American Society of Crime Lab Directors/Laboratory Accreditation Board, International (ASCLD/LAB, International) - this represents an externally enforced professionalism standard. Proficiency testing represents an internally imposed standard of professionalism. Laboratories use it to gauge the performance of crime laboratory personnel and to assess whether they are following industry standards (Burch, Durose, Walsh, and Tiry 2016, 3). We offer two hypotheses:

H3a: The probability of adoption increases in accredited agencies.

H3b: The probability of adoption increases in agencies with proficient workforces.

Forensic crime laboratories do not work in isolation. They interact with a number of private and public actors, including vendors. Many agencies who are interested in efficiency outsource some of their functions (Meier and O'Toole 2009). Laboratories that outsource to external vendors are apt to be privy to enhanced information and are also more likely to use industry-standard best practices. Accordingly, they are more likely to adopt new technologies that increase unit efficiency and effectiveness such as robotics technology (e.g. Jun and Weare 2010; Manoharan 2012). The hypothesis is:

H4: The probability of adoption increases in agencies that outsource.

Finally, we consider the organizational context in which a forensic laboratory operates. Some forensic laboratories work as part of a larger, multiple lab consortia rather than smaller, stand-alone units. There is some support in the technology adoption literature for the proposition that larger organizations tend to adopt new technologies and innovations more frequently than their smaller counterparts (Moon 2002, 429). This could be due to larger organizations having greater capacity, being more sensitive to public pressures (for efficiency), or perhaps even the need to gain leverage to handle internal strife among sub-units (Moon 2002, Holden, Norris and Fletcher 2003; Jun and Weare 2009). The fifth hypothesis is:

H5: The probability of adoption increases in agencies with multiple labs.

**DATA DESCRIPTION**

In 2010, BJS conducted The Census of Publicly Funded Forensic Crime Laboratories (CPFFCL), a survey of the workload and operations of publicly funded crime labs in 2009.[8] The BJS conducted its fourth Census of Publicly Funded Forensic Crime Laboratories on the workload and operations of the nation's crime labs during 2014 and to examine changes since the previous censuses conducted in 2009. [9] The CPFFCL includes all state, county, municipal, and federal crime labs that (1) are solely funded by government or whose parent organization is a government agency and (2) employ at least one full-time natural scientist who examines physical evidence in criminal matters and provides reports and opinion testimony with respect to such physical evidence in courts of law.[10] The CPFFCL does not include operations that engage exclusively in evidence collection and documentation, such as fingerprint recovery and development, crime scene response, and photography. In addition, the census does not collect data on the forensic services performed by police identification units outside of the crime lab and privately-operated crime labs. About half of the crime labs included in the CPFFCL were part of a multi-lab system (two or more physically separate facilities that were overseen by a single organization). The CPFFCL collects information from each lab in multi-lab systems.

---

[8] U.S. Department of Justice, Bureau of Justice Statistics. CENSUS OF PUBLICLY FUNDED FORENSIC CRIME LABORATORIES, 2009. ICPSR34340-v2. Conducted by the University of Illinois at Chicago and Sam Houston State University. Ann Arbor, MI: Inter-University Consortium for Political and Social Research.

[9] U.S. Department of Justice, Bureau of Justice Statistics. CENSUS OF PUBLICLY FUNDED FORENSIC CRIME LABORATORIES, 2014 [Computer file. ICPSR36759-v1. Conducted by the University of Illinois at Chicago and Sam Houston State University. Ann Arbor, MI: Inter-University Consortium for Political and Social Research.

Crime labs employed 14,300 full-time personnel in 2014. The combined operating budgets for the 409 crime labs in 2014 was $1.7 billion. Labs serving state jurisdictions accounted for nearly half of the overall budget in 2014, and labs with 25 or more employees accounted for more than 80 percent of the total combined budget nationwide. (Durose, et al. 2016, 5).

---

[10] In 2010, BJS conducted a third census on the workload and operations of publicly funded crime labs in 2009. A total of 397 of the 411 eligible labs responded to the 2009 census, including at least one from every state. The date(s) of collection were: November 2010 - May 2011. The units of analysis were publicly funded forensic crime laboratories, which was comprised of 411 federal, state, and local forensic crime labs operating during 2009.

In April 2015, the Urban Institute initiated the data collection on behalf of BJS through a web-based data collection interface and mailed questionnaire. Follow-up emails and phone calls were made to nonrespondents and labs that submitted incomplete questionnaires. Of the 409 eligible crime labs that received the questionnaire, 360 (88 percent) provided responses to at least some of the items. Of the 360 respondents, 351 (98 percent) completed the questionnaire through the automated web system, including at least one from every state. The date(s) of collection were: April 2015 - September 2015. The units of analysis were publicly funded forensic crime laboratories, which was comprised of 409 federal, state, and local forensic crime labs operating during 2014.

In 2014, 88 percent of the nation's crime labs were accredited by a professional organization, up from 70 percent in 2002. State crime labs were more likely to be accredited than labs operated by other jurisdictions in 2014. Crime labs employing more full-time employees were more likely to be accredited in 2014. The more forensic functions that a crime lab performs, the more likely it is to be accredited in 2014. Eighty-three percent of crime labs held an international accreditation standard in 2014 (Burch, et al. 2016, 1).

In 2014, 98 percent of crime labs conducted proficiency testing. Nearly all crime labs evaluated the technical competence of employees through declared examinations. The proportion of crime labs conducting blind examinations and random case reanalysis decreased from 2002 to 2014. In 2014, federal crime labs were more likely than county, state, and municipal labs to test the proficiency of employees through blind examinations to conduct random case reanalysis than labs operated by other jurisdictions (Burch, et al. 2016, 4).

In 2014, three-quarters of crime labs had written standards for performance expectations, up from 2009. Written standards establish a threshold for employee performance and ensure that performance measures are applied consistently across employees with similar roles. Federal crime labs were most likely and county labs were least likely to have written standards for performance in 2014. The overwhelming majority of crime labs maintain a written code of ethics. Ethical codes guide behaviors to ensure analysts work within the confines of their expertise, provide objective findings and testimony, avoid conflicts of interest, and avoid susceptibility to outside influences (Burch, et al. 2016, 5).

More crime labs employed externally certified analysts in 2014 than in 2009. Municipal crime labs were most likely to employ at least one externally certified analyst, while federal crime labs were least likely to do so. The larger the number of personnel employed by a crime lab, the likelier that at least one externally certified analyst is on staff. Crime labs that performed eight or more forensic functions were more likely than crime labs performing fewer functions to employ at least one externally certified analyst (Burch, et al. 2016, 5-6).

More than half of federal crime labs dedicated resources to research in 2014. Overall, an estimated 14 percent of crime labs devoted staff, time, supplies, or other resources to forensic science research in 2014. Federal crime labs are far more likely than any other jurisdiction to engage in forensic science research. This research includes experimentation aimed at the discovery and interpretation of facts, revision of accepted methods, or practical application of new or revised methods or technologies (Burch, et al. 2016, 6).

We center our attention on one dependent variable and six independent variables. All descriptive statistics (by year) are located in Table 1. The variable Robotics is a dichotomous dependent variable that is a response to the question of whether the lab uses robotics for any purpose. As the descriptives show, the proportion using robotics increased from 0.41 in 2009 to 0.54 in 2014. While this is a coarse measure, it represents the best publicly-available data for these organizations.

The six independent variables fall into three groupings. The first two - Budget and Requests Received - measure the resources and task environment concepts discussed above. Because both are skewed, they are transformed using a Box-Cox zero-skew transformation.

The second grouping includes the Accreditation and Proficiency variables. Both of these are indices built from multiple underlying indicators. In the case of Accreditation, we measured the number of accreditations given the responses to four questions, each about whether the lab had received a specific form of accreditation (two types from the American Society of Crime Laboratory Directors, one from Forensic Quality Services, and any other). This variable ranges from zero to three in both years but it is possible that a lab could have agreed to all four questions. In the case of Proficiency, we also built an index from four questions asking whether staff were tested in four different ways (by blind test, by declared examination, by random case reanalysis, or by any other). In 2009, some labs tested proficiency by all four ways, although in both years some labs did no proficiency testing at all.

The third grouping of independent variables includes two dichotomous indicators. One measures whether labs responded being part of a multiple lab arrangement; the other measures whether they reported outsourcing at all.

Because this is a parsimonious listing of variables, we discuss how our estimation strategy addresses unobservable heterogeneity below.

[Insert Table 1 about here.]

## ESTIMATION AND RESULTS

We fit the models presented below by probit with robust standard errors clustered by state. This helps address unobservable heterogeneity that occurs at the level of the local geography. In the model for 2009, there are 51 clusters because one lab is located in Washington, DC; in the 2014 sample, there are 50 clusters. Both models shown in Table 2 fit the data fairly well, with the Wald test rejecting at a high level of significance the null that all of the coefficients could be restricted to be equal to zero. The pseudo-$R^2$ is lower than hoped for but we also know that the models are fairly parsimonious. We excluded all federal labs from our analysis.

[Insert Table 2 about here.]

Both models show that the probability of adoption is greater in agencies with larger budgets. Both effects are significant at conventional levels. Figure 1a shows the marginal effect of budget for 2009; Figure 1b shows that marginal effect for 2014. While both coefficients are positive and significant, perhaps the most notable differences across time are that the slope is attenuated in 2014, and uncertainty about that effect increases (the width of the confidence interval is greater is 2014). An interpretation is that the effect is less important over time - perhaps suggesting that budget is important for early adoption but less important for later adoption.

[Insert Figures 1a and 1b about here.]

As with budget, both the 2009 and 2014 models show that the probability of adoption is greater in crime labs that receive more requests. The number of requests received serves as a proxy for the agency's task environment. Both positive effects are significantly different from zero at conventional levels. For task environment, Figures 2a and 2b show the marginal effects for 2009 and 2014, respectively. The differences between the marginal effects are slight, although it appears that there is a difference in terms of both slope and uncertainty about the estimated effect. In contrast to the finding about budget, it appears that the effect strengthens over time - that is, that in 2009 the effect of requests received is smaller than in 2014. This finding suggests that task environment is perhaps less important for early adoption than for later adoption of an emerging technology like robotics.

[Insert Figures 2a and 2b about here.]

The other primary finding in both models is that accreditation of crime labs is positively associated with probability of adoption. For both the 2009 and 2014 data, the models show that the probability of adopting robotics increases for agencies as the number of different accreditations increases. Both estimated effects are significantly different from zero at conventional levels. However, the models suggest that we are more certain about the estimated effect for 2014 than 2009. The estimated marginal effects shown in Figures 3a and 3b also suggest that the impact of accreditation is more attenuated than that for task environment in 2014 or budget in 2009 (the estimated slopes are flatter).

[Insert Figures 3a and 3b about here.]

Of the other variables included in the models, only the effect of outsourcing in 2014 is estimated to be statistically significant from zero. That sign is positive, and the impact is moderate. In this 2014 data, the estimated probability of adoption of robotics is 0.47 when crime labs did not outsource, compared to about 0.63 for those that did outsource (all other things being held equal). This positive effect is fairly moderate in terms of impact compared to the findings above. None of the other coefficients in the models are significantly different from zero at conventional levels.

**CONCLUSION**

The purpose of this paper is to demonstrate the extent of adoption of an advanced "smart technology" in the form of robotics in a class of important public agencies (the crime labs). It also provides insight into some of the potential drivers of adoption - in this case, very traditional public administration concerns like budget, the task environment, and the extent of agency professionalism. Combined, the magnitude and statistical significance of the effects indicate that adoption is associated with factors that are common and observable to those interested in prognosticating about the future of smart technology in government.

Moreover, the shape and magnitude of the effects give us some ability to talk about early and late adoption. In this context, having two census surveys (one in 2009 and one in 2014) lets us see how the incidence of robotics changes and also how the variables contribute differently to those likelihoods over time. As we noted previously, it is important to take such evidence as indicative at best and follow up with more in-depth analyses (either through causal identification or maybe even better, in-depth directed focus group inquiries). Yet, our suggestive results paint a picture in which early and late adoption of a smart technology like robotics shifts in response to different drivers.

Two themes warrant special emphasis here, though. First, we know that robotics as an indicator of the fourth industrial revolution or a second machine age may be misguided. The reason for this is simple. If we believe that the next phase of human existence is ahead of us, the robots are already here. Over half of the crime labs used robotics - five years ago. It is expected that the next census will show even greater penetration, but if so should we continue to treat robotics as novel? Either the dating of the next phase is wrong or robotics are the leading edge. The most likely answer is that robotics are probably an established technology and not a truly novel one.

The second theme is more telling. Crime labs are a low-visibility environment. In contrast to Amazon's warehouses or the Adidas "SpeedFactory", government agencies (and especially so government crime labs) are so yesterday. In the narrative, the labs are the domain of bureaucrats in white lab coats protecting their expertise-driven domain and maximizing the perquisites of the job. Yet, the incidence of robotics - and its associated potential drivers - runs against that narrative. Instead of dragging behind, these agencies are pushing ahead. The crux of the matter remains the incidence of such smart technologies. In Gibson's words, the robots are here but not everywhere (yet). But then, the relative incidence of an innovation is an old and continuing story in studies of the public sector, and the relative incidence of smart technologies is a worthy topic for future public administration studies.

**TABLE 1. Descriptive statistics.**

For 2009:

| Variable | Mean | SD | Minimum | Maximum |
|---|---|---|---|---|
| Robotics | 0.413 | 0.493 | 0 | 1 |
| Budget | 27.358 | 4.854 | 11.241 | 38.536 |
| Requests Received | 15.355 | 4.177 | 4.095 | 31.427 |
| Accreditation | 0.892 | 0.546 | 0 | 3 |
| Proficiency | 1.476 | 0.766 | 0 | 4 |
| Multiple labs | 0.394 | 0.490 | 0 | 1 |
| Outsourced | 0.297 | 0.458 | 0 | 1 |

For 2014:

| Variable | Mean | SD | Minimum | Maximum |
|---|---|---|---|---|
| Robotics | 0.535 | 0.500 | 0 | 1 |
| Budget | 46.675 | 12.701 | 15.208 | 68.863 |
| Requests Received | 12.501 | 2.482 | 4.773 | 20.340 |
| Accreditation | 0.996 | 0.443 | 0 | 3 |
| Proficiency | 0.454 | 0.612 | 0 | 3 |
| Multiple labs | 0.498 | 0.501 | 0 | 1 |
| Outsourced | 0.373 | 0.484 | 0 | 1 |

**Table 2. Probit models.**

|  | 2009 Dataset | | | 2014 Dataset | | |
|---|---|---|---|---|---|---|
| Variable | Coefficient | Robust SE | | Coefficient | Robust SE | |
| Budget | 0.121 | 0.032 | *** | 0.016 | 0.007 | ** |
| Task Environment | 0.073 | 0.034 | ** | 0.201 | 0.032 | *** |
| Accreditation | 0.341 | 0.186 | * | 0.541 | 0.196 | *** |
| Proficiency | -0.067 | 0.114 | | 0.120 | 0.131 | |
| Multiple Labs | -0.009 | 0.213 | | -0.105 | 0.181 | |
| Outsourcing | 0.094 | 0.162 | | 0.515 | 0.156 | *** |
| Constant | -4.980 | 0.787 | *** | -3.906 | 0.478 | *** |
| | | | | | | |
| Wald $\chi^2$ | 50.00 | | | 97.72 | | |
| Pseudo $R^2$ | 0.26 | | | 0.21 | | |
| N | 269 | | | 271 | | |

All models include robust standard errors clustered by state.

* indicates significance better than 0.10 (two-tailed test)
** indicates significance better than 0.05 (two-tailed test)
*** indicates significance better than 0.01 (two-tailed test)

**Figure 1a. The marginal effect of budgets (2009 data).**

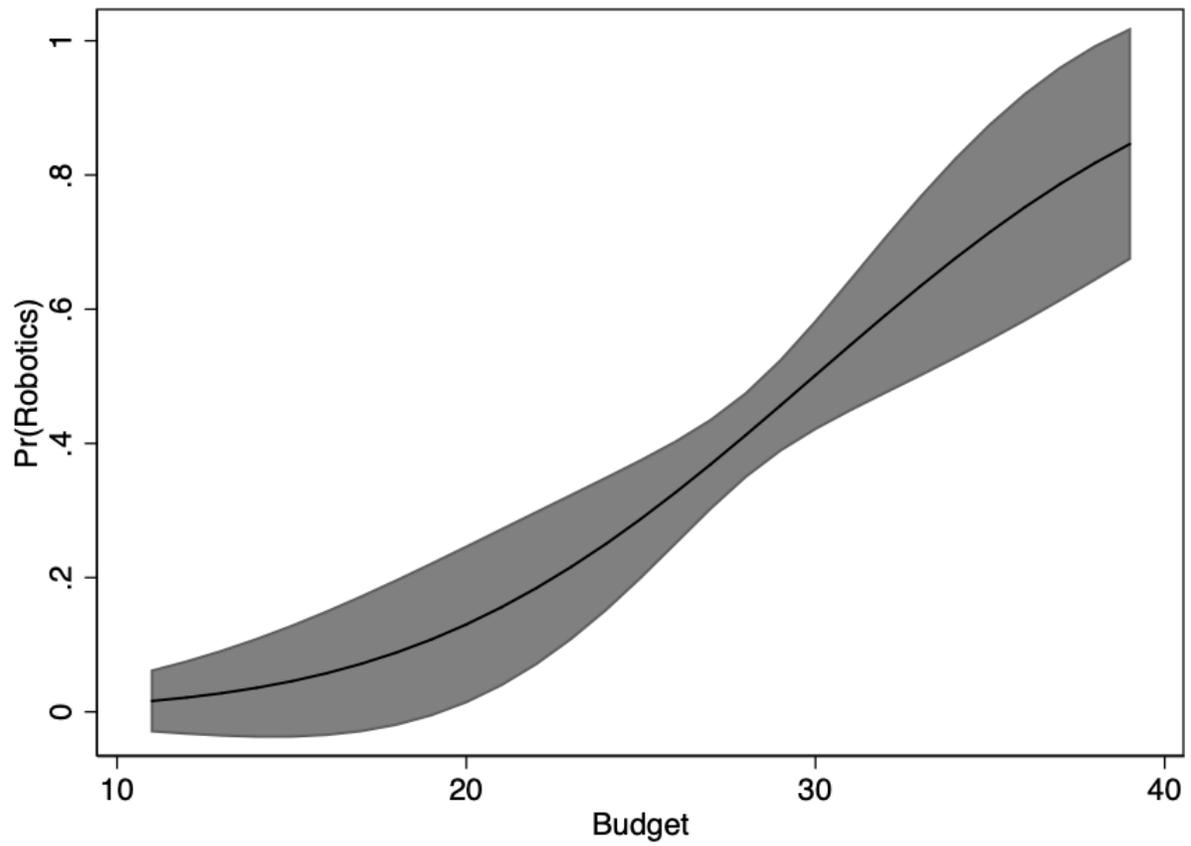

**Figure 1b. The marginal effect of budgets (2014 data).**

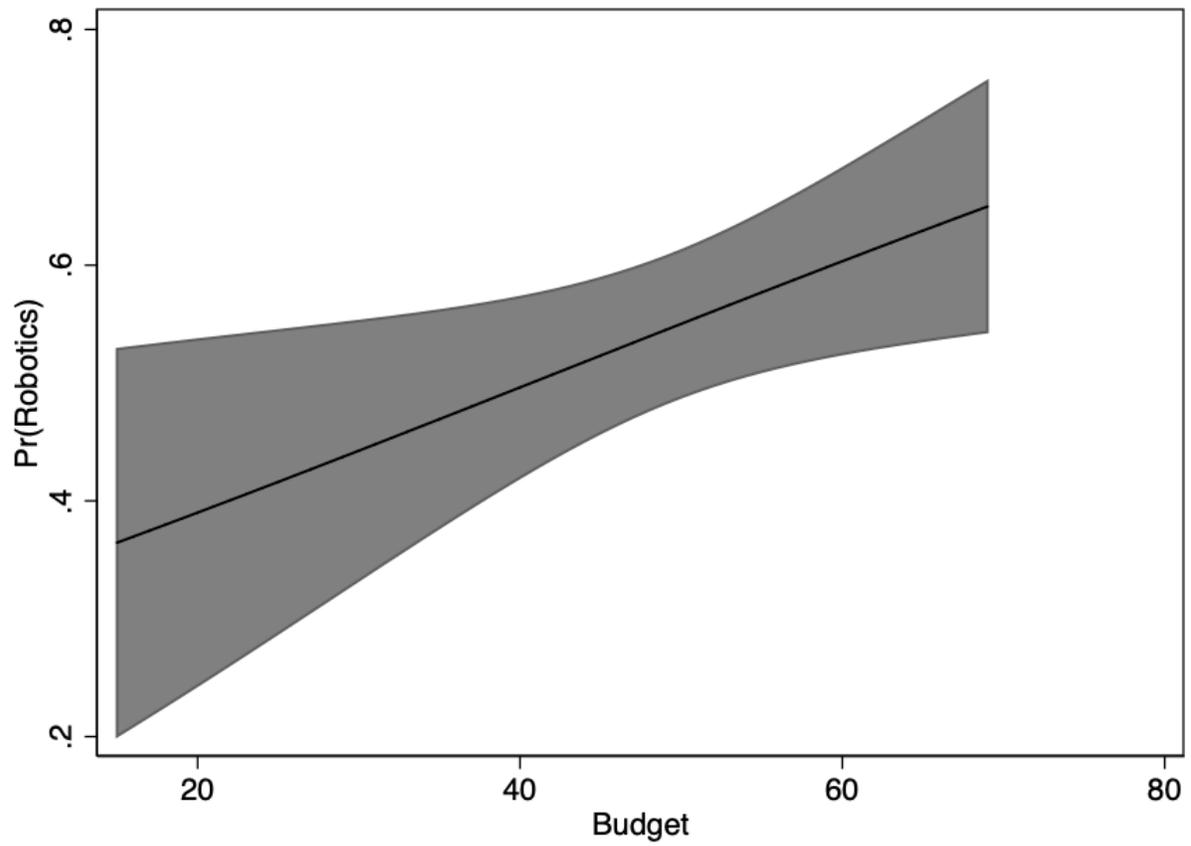

**Figure 2a. The marginal effect of task environment (2009 data).**

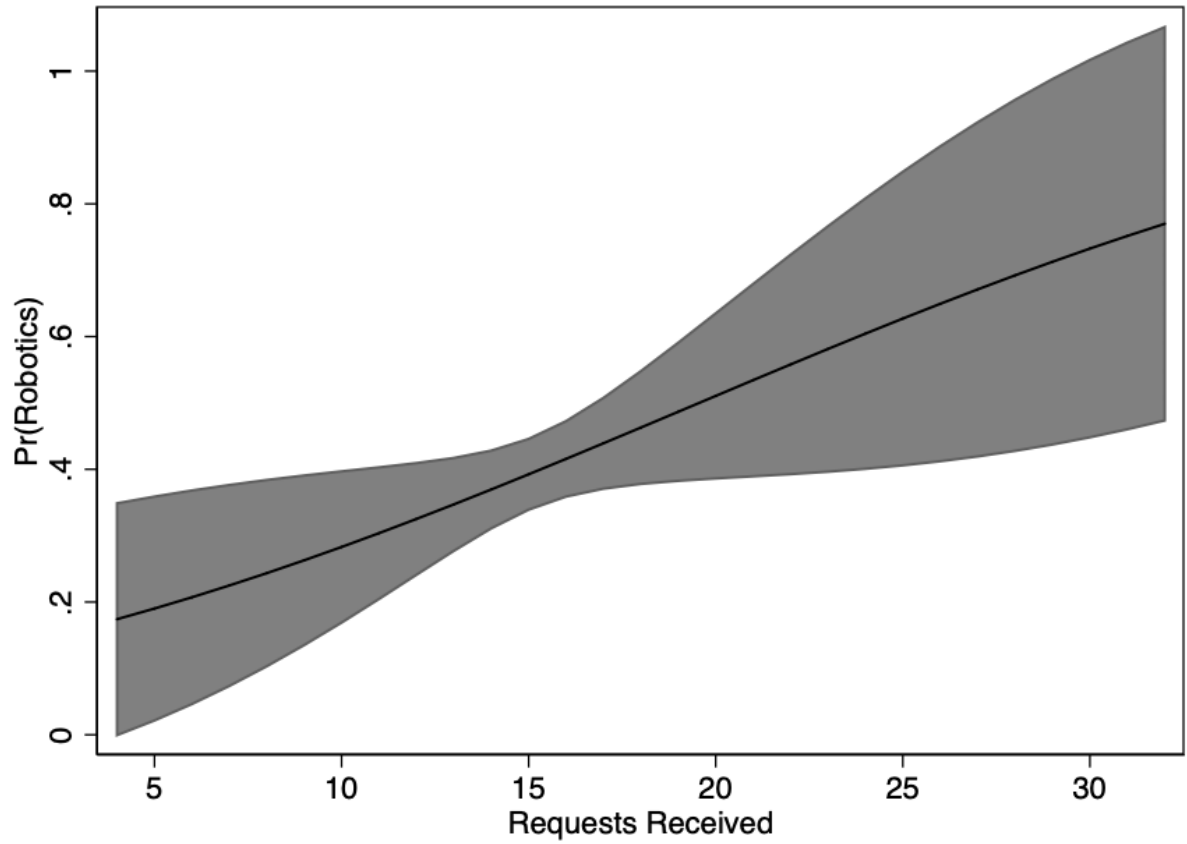

**Figure 2b. The marginal effect of task environment (2014 data).**

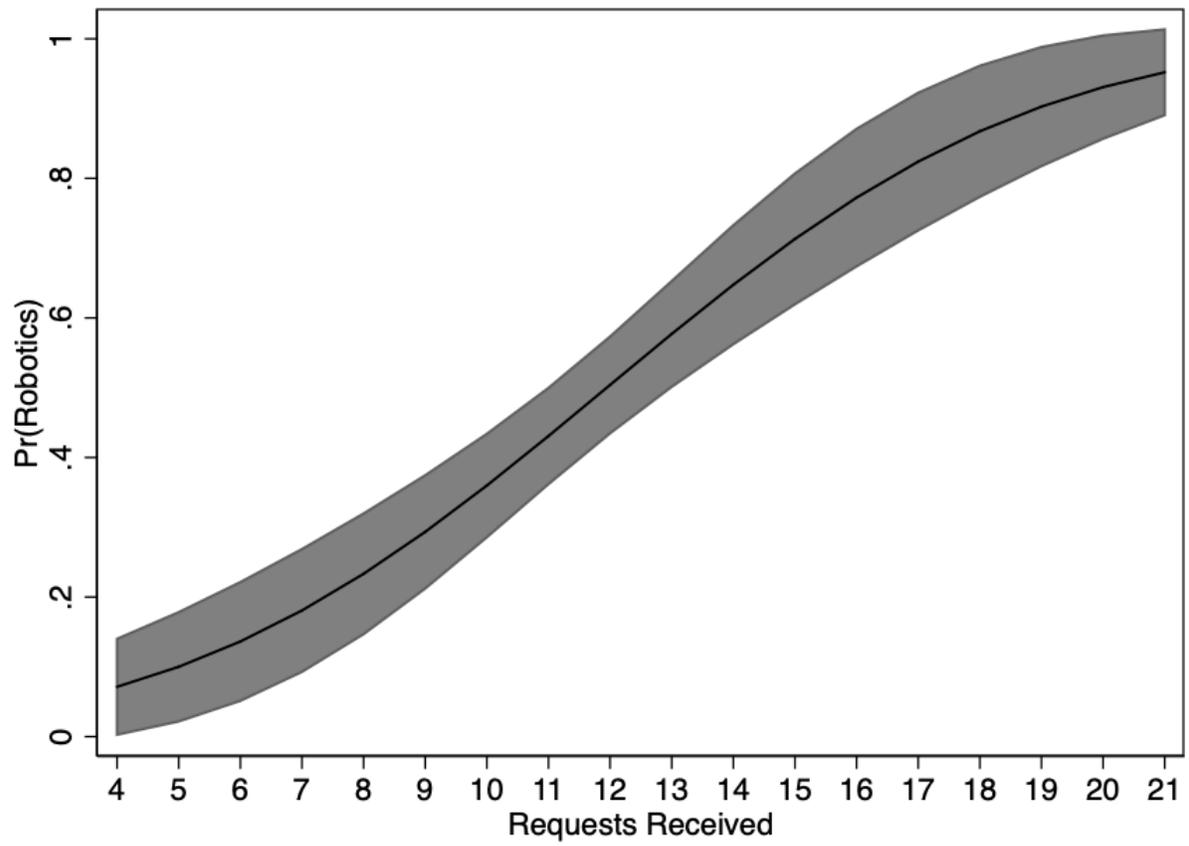

**Figure 3a. The marginal effect of professional accreditation (2009 data).**

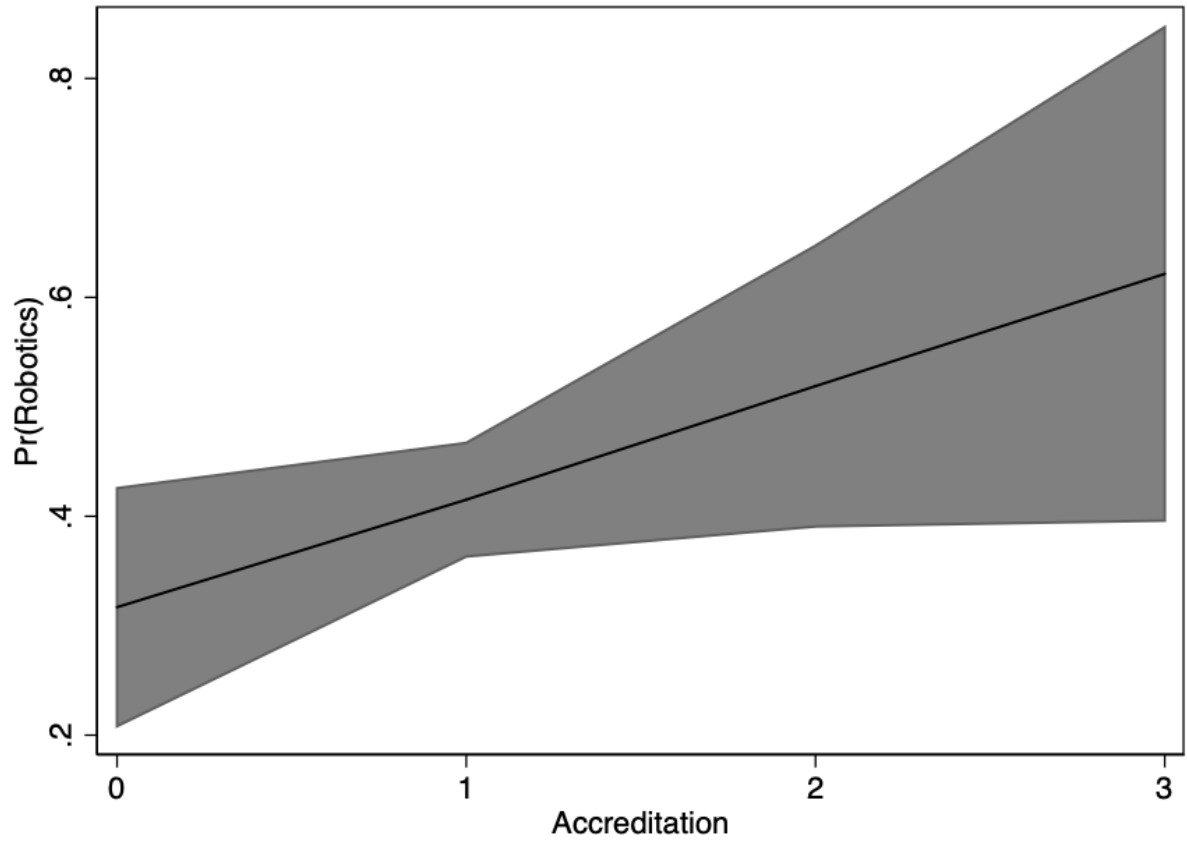

**Figure 3b. The marginal effect of professional accreditation (2014 data).**

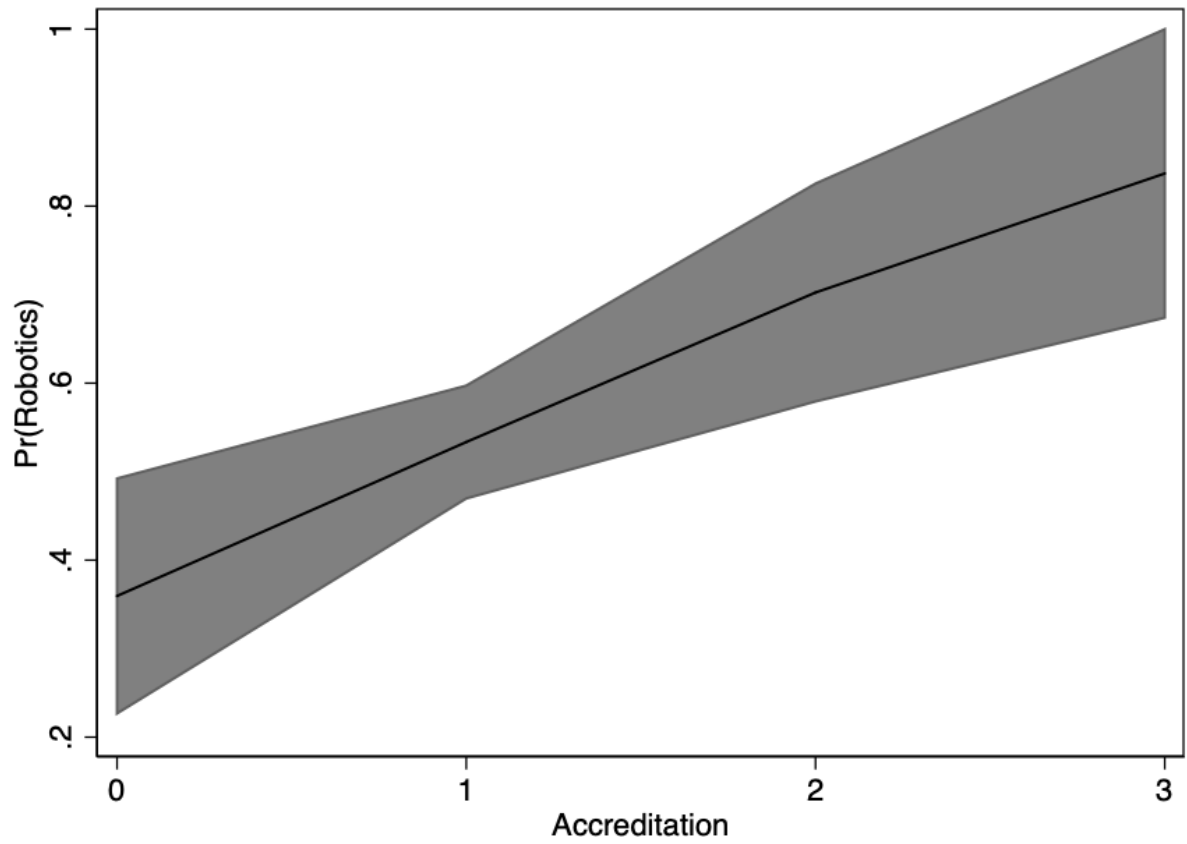